\documentclass[runningheads]{llncs}
\usepackage{pgfplots}
\pgfplotsset{compat=1.18} 

\usepackage{enumitem}

\usepackage{adjustbox}
\usepackage{float}
\usepackage{multirow}
\usepackage[section]{placeins}
\usepackage{pgf-pie}  
\usepackage{array}
\usepackage{xurl}

\definecolor{ibmyellow}{HTML}{ffb000}
\definecolor{ibmorange}{HTML}{fe6100}
\definecolor{ibmmagenta}{HTML}{dc267f}
\definecolor{ibmindigo}{HTML}{785ef0}
\definecolor{ibmblue}{HTML}{648fff}

\usepackage{hyperref}
\hypersetup{
    colorlinks=true,
    linkcolor=ibmmagenta,
    filecolor=ibmorange,      
    urlcolor=ibmblue,
    citecolor=ibmmagenta,
    }

\usetikzlibrary{cd}

\usepackage[T1]{fontenc}
\usepackage{graphicx}
\usepackage{color}

\begin{document}
\title{Defining DLT Immutability: A Qualitative Survey of Node Operators}
%
%
\author{
Alex Lynham\inst{1}\orcidID{0009-0005-0488-7651} 
\and
Geoff Goodell\inst{1}
}
\authorrunning{A. Lynham and G. Goodell}
%
\institute{$^1$University College London, 66-72 Gower St, London WC1E 6BT, England}

\maketitle              

\setcounter{footnote}{1}

\begin{abstract}

Immutability is a core design goal of permissionless public blockchain systems. However, rewrites are more common than is normally understood, and the risk of rewrite, cyberattack, exploit or black swan event is also high. Taking the position that strict immutability is neither possible on these networks nor the observed reality, this paper uses thematic analysis of node operator interviews to examine the limits of immutability in light of rewrite events. The end result is a qualitative definition of the conditional immutability found on these networks, which we call \textit{Practical Immutability}. This is immutability contingent on the legitimate governance demands of the network, where network stakeholders place their trust in the governance topology of a network to lend it legitimacy, and thus manage ledger state.

\keywords{blockchain \and decentralization \and DAO \and governance \and security}
\end{abstract}

\newcommand{\blockquote}{\medskip \noindent \leftskip 16pt}

\newcommand{\blockquotesmallskip}{\smallskip \noindent \leftskip 16pt}

\newcommand{\blockquoteendsmallskip}{\smallskip \leftskip 0pt}

\newcommand{\blockquoteend}{\medskip \leftskip 0pt}

\newcommand{\blockquoteendnoskip}{\leftskip 0pt}


   


\section{Introduction}

\begin{figure*}[ht]
\centering
\begin{tikzpicture}[
    captiontext/.style={below=3mm, text width=5cm}
]

\pie[
    color = {
        ibmmagenta,
    ibmyellow,
    ibmblue}, 
  sum = auto,
    radius = 1.8,
    rotate = 270,
    text=legend
]
    {9/Undecided,
  10/No,
  10/Yes}
\node [captiontext] at (current bounding box.south) {Q9. Would [your team] knowingly run code that caused a ledger rewrite that affected a single account balance?};
\end{tikzpicture}
\begin{tikzpicture}[
    captiontext/.style={below=3mm, text width=5cm}
]

\pie[
    color = {
    ibmyellow,
    ibmblue}, 
  sum = auto,
    radius = 1.8,
    rotate = 270,
    text=legend
    ]
{
    13/No,
  16/Yes}
\node [captiontext] at (current bounding box.south) {Q10. Have [your team] knowingly run code that caused a ledger rewrite that affected a single account balance?};
\end{tikzpicture}
\begin{tikzpicture}[
    captiontext/.style={below=3mm, text width=5cm}
]

\pie[
    color = {
    ibmyellow,
    ibmblue}, 
  sum = auto,
    radius = 1.8,
    rotate = 270,
    text=legend
]{
    14/No,
  15/Yes}
\node [captiontext] at (current bounding box.south) {Q11. Have [your team] knowingly run code that caused a ledger rewrite that affected multiple account balances?};
\end{tikzpicture}
\caption{Validator participation in rewrite events}
\label{fig:q9}
\end{figure*}
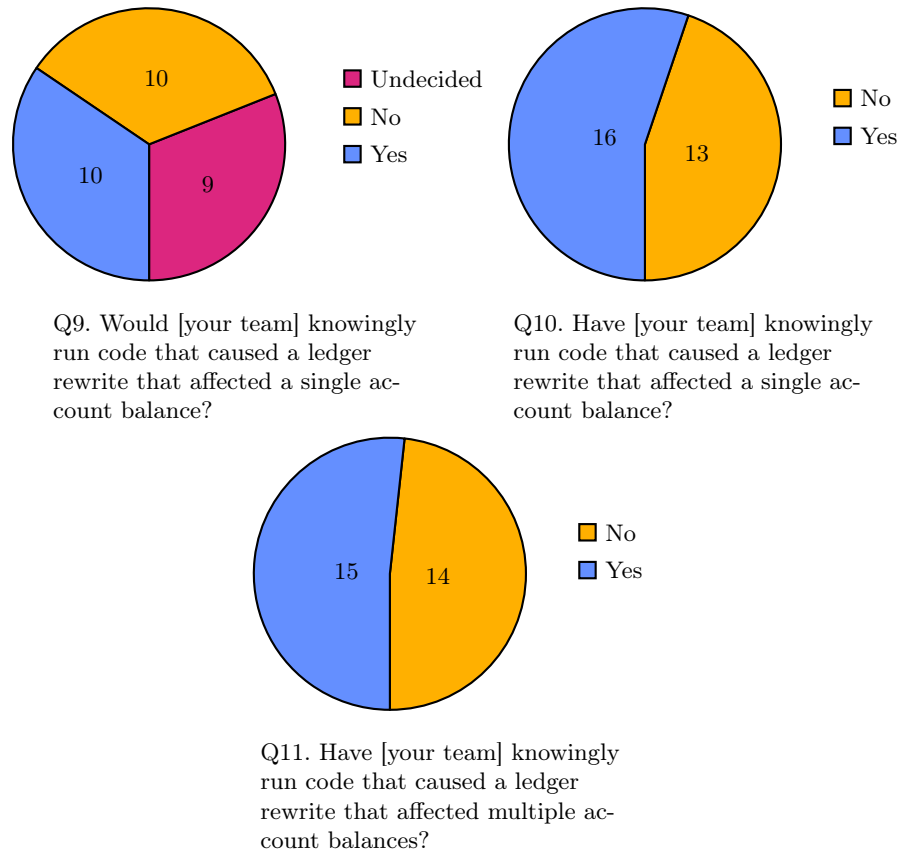

In this paper, we make the claim that immutability is impossible to guarantee on permissionless public blockchains, and attempt to find a qualitative definition that more closely matches network participants' experience. To do this, we make use of interviews and surveys with blockchain node operators from the Cosmos Ecosystem,\footnote{The Cosmos Ecosystem is an interconnected ecosystem of blockchains built using the Cosmos Software Development Kit (SDK). They are linked by the IBC protocol, and are Delegated Proof-of-Stake by default, built around a DAO (Decentralized Autonomous Organisation) for voting on protocol upgrades and governance.~\cite{whatiscosmos}} which has experienced a number of high-profile rewrite events. While some operators do not hold firm to the principle of immutability, at least half of surveyed operators still take the position that blockchains should be immutable, (see Fig \ref{fig:q9}) a position generally held by blockchain participants, regulators, and the wider public. The research question this paper addresses is:

\vspace{0.5em}\blockquotesmallskip
\textit{If ledger immutability is not absolute, then under what conditions is it considered adequately immutable by stakeholders?}

\vspace{0.5em}\blockquoteendsmallskip

In Section 2, we describe our methodology, discuss existing analysis of immutability, and outline our desire to coin a working definition for a \textit{conditional} immutability we call \textit{Practical Immutability}. Crucially, unlike other analyses, we see the trust relationship as being conditional on the ability to \textit{change} the ledger in certain situations, rather than finding that trust is conditional on immutability of the ledger data. In Section 3 we examine node operators' opinions on governance maturity and software upgrades, two potential vectors for ledger state changes, as well as their views on participating in ledger rewrite events. We then use thematic analysis to aggregate themes into an analysis of the limits of real-world immutability on public, permissionless networks in Section 4.

The meta-themes we have identified are \textit{Legitimacy} and \textit{Trust}, which respectively govern the correctness of ledger state and the authority of governance being seen as legitimate, if not non-arbitrary. We theorize that, as intuition would dictate, ledger state can be mutated subject to the consent of what is perceived as a legitimate governance process. Although blockchains are typically viewed in the context of rational agents, it is a mixture of calculative and non-calculative trust that supports a network's governance topology and lends it its legitimacy.

This leads to the conclusion that these ledgers were not designed to be immutable in the strictest definition, though it seems likely that this is an emergent reality rather than how these blockchains were designed to function. As the saying goes, it is a feature, not a bug. In operation, their immutability protects against opportunistic attacks by hostile agents, but not exogenous attacks (such as social engineering), or against social attacks, co-ordinated user actions or the governance decisions of the network. For most network participants, this is an acceptable threshold to maintain their trust.

\section{Methodology and Context}

\subsection{Methodology}

\begin{table}[t!]
    \begin{adjustbox}{width={0.98\textwidth},totalheight={\textheight},keepaspectratio}
\begin{minipage}[t]{.65\textwidth}
{\renewcommand{\arraystretch}{1.5}
\quad{\setlength{\tabcolsep}{0.4em}
\centering
    \caption{Survey and semi-structured interview recruitment \strut}
    \label{table:recruitment}
    \begin{tabular}{| p{0.25\linewidth} | p{0.25\linewidth} | p{0.25\linewidth} | p{0.25\linewidth} |}
         \hline
          \raggedright\textbf{Artifact}\arraybackslash& \raggedright\textbf{Number of recruitment forums}\arraybackslash& \raggedright\textbf{Estimated Reach}\arraybackslash& \raggedright\textbf{Participants}\arraybackslash \\
         \hline
         Survey&   5&500& 35\\
         \hline
         Semi-structured interviews&   6&550& 7\\
         \hline
    \end{tabular}
}}
\end{minipage}%
\qquad\qquad
\begin{minipage}[t]{.65\textwidth}
{\renewcommand{\arraystretch}{1.5}
\quad{\setlength{\tabcolsep}{0.4em}
\centering
    \caption{Number of chains validated by node operators' organisations \strut}
    \label{table:chains_validated}
    \begin{tabular}{| p{1\linewidth} | }
         \hline
          \textbf{Number of Chains validated} \\
         \hline
         About 15 \\
         \hline
         More than 20 \\
         \hline
         83 \\
         \hline
         About 50 \\
         \hline
         About 100 \\
         \hline
         About 20 \\
         \hline
    \end{tabular}
}}
\end{minipage}
    \end{adjustbox}
\end{table}
We decided to conduct a survey and semi-structured interviews of node operators. These practitioners run, or work for organisations that run validator nodes. Although it might have been possible to do a wider survey of stakeholders, we decided that technical practitioners would have a more frank and low-level view of the operation of these protocols and incentive mechanisms. For example, protocol engineers design systems in an idealistic world, while node operators often find out the limitations of these systems at runtime, in the form of bugs, zero-days and design limitations.

\subsection{Recruitment and protocol}
To find these operators, we posted an anonymous survey in private validator communication channels, private forum threads and network communication channels in the Cosmos Ecosystem. These are typically kept up-to-date by validators, blockchain Foundations and Core Teams, with only active validators being able to view them for cybersecurity reasons. These channels, as well as additional channels to which the researchers had access\footnote{This was because the survey had Cosmos-specific questions, and the semi-structured interviews could be more generic.} were then used for recruiting interview participants. The full survey responses are in our companion paper,~\cite{lynham2025decentralizationqualitativesurveynode} from which we have reproduced two relevant figures in the Appendix of this paper.

In light of reputational and personal risks involved in participating in this study, it could only proceed under a strict protocol of anonymity for participants. This means that we do not have demographic information or even role portraits for participants. Nevertheless, we do have some estimated figures for recruitment (Table \ref{table:recruitment}) which we reproduce here from our companion paper. The interview began with a script before an agreement to begin recording. The first question was an ice-breaker, asking the number of chains validated by the node operator's organisation. This optional question was intended to not only begin the discussion, but also validate the operator's claim as an operator (not every interviewee gave an answer; nevertheless, the data are reproduced in Table \ref{table:chains_validated}). The same discussion areas were used for each interview, and the interviewer prompted for clarifications were they were needed. After being transcribed the source audio was deleted and the transcripts were then identified by a UUID.

\subsection{Immutability}

A good starting definition for immutability is the ISO one, (ISO 22739:2024 3.51); the ``property of a distributed ledger (3.23) wherein ledger records (3.55) cannot be modified or removed once added to that distributed ledger.''~\cite{isoblockchain} However, in light of the argument that these systems are equally, if not more, reliant on social consensus than technical means to secure finality, we conclude that such a rigid definition is sufficient for theoretical analysis of DLT systems, but not for analysis of the real-world systems in which they are situated, nor the prospective design of future systems that address similar objectives. 

As early as 2018, Kim and Wang argued that an ``immutability measure'' was needed to indicate ``the degree of difficulty in mofidying existing data in Blockchain,'' in light of ``the probability of of attackers succeeding in modifying legitimate data in [blocks].''~\cite{kim_wang_immutability_measure} As in the ISO definition of decentralization, they make a clear distinction between decentralization and immutability, characterizing decentralization in terms of dispersion of authority: ``The key properties of blockchain are immutability of data and decentralized authority based on peer-to-peer networks.''

Prasad et al. took this further, analysed Bitcoin, Ethereum and Ethereum Classic (BTC, ETH and ETC, respectively) and concluded that ``there are various features of immutability in blockchain networks that can still be broken on the social level. Even if Nakamoto Consensus is still valid, certain social coordination by network participants can break aspects of blockchain immutability.''~\cite{quantifying_blockchain_immutability_over_time} They conceptualise blockchain immutability not in a binary way at network level, but as eight subsystems,\footnote{Monetary Policy, Consensus Mechanism, Hashing Algorithm, Chain State, Reserved State Space, Transaction History, Transaction Finality, and Social Contracts.} which are either mutable (0) or immutable (1).\footnote{It is important to state that at this point they are in practice all mutable, due to the exact social attacks they describe. Our argument is that on a long enough time scale, these are all zero, and only guaranteed by their continued perception as legitimate by governance. Indeed, even a technically difficult change, for example to the consensus mechanism, can happen---for example Ethereum moving to Proof-of-Stake, which occurred after their paper was written.}

\subsection{Practical Immutability}
In the previous definitions, the relationship is commonly that immutability of data is a prerequisite for trust. As we will argue in Section 4.1, this relationship is inverted when agents require that state to be subject to reversion or change at the behest of governance. 

The goal of this paper to arrive at a definition of immutability, \textit{Practical Immutability}, that can be used for analysing heterogenous DLTs. We are not the first to note that there is a paradox in the use of the term ``immutability.'' Walch observes that immutability is a ``defining'' feature of blockchains,\footnote{At the time of writing, Walch notes that the most-cited blockchain paper on SSRN explicitly characterized blockchains in terms of ``immutability,''~\cite{blockchain_technology_principles} which is still the case in more recent surveys such as Bashir.~\cite{Bashir2022} } while suggesting that most practitioners mean ``hard to change,'' not ``unchangeable'' by ``immutable.''~\cite{walch2017} Elsewhere, Lemieux argues ``immutability is best viewed not as a property of blockchain-based ledgers but as a sustained commitment that a group of individuals holds onto because they believe that the attribute is desirable, even necessary.''~\cite{Lemieux_2022} Whether due to ideological reasons or professionalization (see Section 4.2) this matches our analysis; if immutability is not seen as desirable, it is no longer ``necessary.''

In our companion paper,~\cite{lynham2025decentralizationqualitativesurveynode} we argued that differences in the governance topology and network topology of a network simply change the manifestation of ledger rewrites, not the rewrite itself. Adding to this zero-day bugs, black swan events and economic and social attacks from within the governance topology of ledgers,\footnote{It should be noted the most famous instance of a social attack that resulted in a rewrite via a ledger split (ISO 22739:2024 3.56), generally referred to as practitioners in our fieldwork as a `hard fork,' is the rewrite that split Ethereum into Ethereum and Ethereum Classic after the DAO hack. Indeed, whether it was an attack at all is something of a fraught topic, which shows the motivation for this paper.} it is not possible to guarantee immutability on public permissionless ledgers. This begs the question, when practitioners talk about `immutability', what exactly is it that they mean? This is the question we hope to answer in this paper.

In systems like (Delegated) Proof-of-Stake blockchains that rely on economic security to secure finality, Budish et al., have argued that attacks are only punished after-the-fact, by the collapse of the endogenous token of these ledgers. Thus, these attacks are only made ``expensive'' to the attacker by the ``collapse'' of the token, and are ``cheap'' otherwise.~\cite{budish2024economiclimitspermissionlessconsensus} In systems where an attack is only ``expensive due to collapse'' there is not an ex-ante mechanism that guarantees the integrity of the ledger. This is an old problem; in his Bitcoin whitepaper, Satoshi Nakamoto describes the same collapse mechanic, ``If a greedy attacker is able to assemble more CPU power than all the honest nodes, he would have to choose between using it to defraud people by stealing back his payments, or using it to generate new coins. He ought to find it more profitable to play by the rules, such rules that favour him with more new coins than everyone else combined, than to undermine the system and the validity of his own wealth.''~\cite{bitcoin_nakamoto}

\section{Results}
What follows are summary insights from our semi-structured interviews. The discussion areas this paper focusses on are, ``How often do you or your node operations team check the code contained in an upgrade before applying it?'', ``Have you or your node operations team knowingly run code that caused a ledger rewrite that affected a single account balance?'' and ``What is your opinion on the level of maturity in blockchain governance?''

Specific examples, concepts and real-world scenarios from the transcripts were coded, with some examples and concepts being coded into multiple categories. These were then grouped into higher-level themes (this process is documented in Table \ref{table:coding} in Appendix F, but omitted for brevity and presentation reasons in Tables \ref{table:results_coding1} - \ref{table:results_coding4}). This process was inductive initially, and then aggregation occurred via situating the concrete examples within the context of the wider transcripts and existing literature on decentralization and ledger immutability.~\cite{lynham2025decentralizationqualitativesurveynode} The full transcripts of discussions on the topic areas of validator due diligence on code upgrades, rewrite events and governance maturity from the semi-structured interviews and survey responses can be found in the Appendix.

These discussion areas were chosen to analyse immutability for a number of reasons. The upgrade of a chain, especially a hard-fork, is often when a state change occurs, since it is simpler to manipulate the state of a halted chain and then if necessary update any invariant checks in the software before nodes restart. A ledger rewrite affecting a single account balance, meanwhile, is the single most targeted type of rewrite, and the most controversial. Finally, the question of governance maturity was selected as it gives further insight into the trust dynamics at play in the on-chain and off-chain interaction of stakeholders.

Across these discussion areas, immutability was often discussed; perhaps surprisingly, many operators volunteered a flexible approach to ledger mutability. For some, this was pragmatism in the face of changing circumstances; for some, adherence to the wishes of stakers voting; for others, risk management. Accordingly, governance was mentioned frequently, often hand-in-hand with any discussion of rewrites. Incentives were discussed in terms of validator responsibilities, particularly around voting and upgrade integrity, and trust dynamics and (de)centralization were recurring themes as the context throughout. Wherever governance was mentioned, trust and legitimacy were usually implicit to the framing of any opinions over strengths or weaknesses, or discussions of specific events related to chain governance.

We found that ledger state is often treated by operators---those trusted explicitly with its management---as something that is less than absolutely final. Stepping above the themes of conditional immutability, state rewrite events, trust dynamics, incentive function, and governance, there are identifiable aggregate themes, which could be summarized as validators' recognition (for better or worse) of supremacy of governance over the technical topology on the one hand,\footnote{Indeed, one validator went so far as to describe their perspective as ``nihilistic'' (Appendix B, Response 6) as a result of this reality.} and their acceptance of the importance of trust on the other. 

A good model for this trust is that described by Williamson, where calculative trust is what a layperson might describe as calculated risk.\footnote{Williamson is keen to stress that trust and risk are not interchangeable, and critiques the idea that taking on risk implies trust.} Elsewhere, it may be that a combination of interpersonal trust---which is typically non-calculative---and institutional trust means that trust dynamics in these networks are less calculative than might be ordinarily assumed, ``[i]nstitutional trust refers to the social and organizational context within which contracts are embedded. In the degree to which the relevant institutional features are exogenous, institutional trust has the appearance of being non-calculative. In fact, however, transactions are always organized (governed) with reference to the institutional context (environment) of which they are a part. Calculativeness thus always reappears.''~\cite{williamson1993} Williamson assumes a high bar for personal or non-calculative relations; we do not.\footnote{For instance, he excludes all commercial relationships from personal trust, which seems to downplay the extent to which irrationality can drive non-calculative behaviour.} This combination of calculative and non-calculative interaction is something we described at length in our discussion of the typical agents in a blockchain network.~\cite{lynham2025decentralizationqualitativesurveynode} Even those validators that accepted it was their responsibility to ensure the integrity of the ledger saw this as a challenging prospect in light of operational realities, so if there is a tension between a pragmatic attitude towards rewrites versus immutability, it is largely implicit rather than explicit.

{\renewcommand{\arraystretch}{1.5}
\quad{\setlength{\tabcolsep}{1em}
\begin{table}[t!]
    \begin{adjustbox}{width={\textwidth},totalheight={\textheight},keepaspectratio}
    \begin{minipage}[t]{.62\textwidth}
    \caption{Thematic analysis (a): 1 of 4}
    \label{table:results_coding1}
    \begin{tabular}{|p{0.8\linewidth}|>{\raggedright\arraybackslash}p{0.2\linewidth}|}
         \hline
         \raggedright \textbf{Concept or discrete example}\arraybackslash & \textbf{Theme} \\

         \hline
         ``Whatever has happened on-chain, has happened, and it should not be changed afterward.'' (Appendix B, Response 1)&  Conditional Immutability\\

         ``We never take an absolutist approach\ldots In principle, blockchain is supposed to be immutable, but in some cases we carve out exceptions.'' (Appendix B, Response 2) & \\
         ``I think that if everybody just thinks ``the validators are checking [the code],'' that's a very dangerous place to be, because I think it's very rare that people are checking it to at least the level that it would need to be, in order to be confident about it.'' (Appendix C, Response 4) & \\

         \hline
         ``Untombstoning is cases when a validator has double-signed, for example, during an upgrade, and then they make a proposal to [get reinstated]\ldots since this is a governance-based change, when the majority clams that this is fine, then basically everybody has to follow suit. In my opinion this should not be done.'' (Appendix B, Response 1) &  Rewrite Events \\

         ``We think that it's an immutable ledger, but sometimes, shit happens, sometimes [there is a rewrite] for a good reason\ldots The community has a discussion about it and then different people make up their mind, and then [as a validator] our responsibility, once it's passed, we just upgrade the chain, right? If you disagree, you can always exit. It's permissionless.'' (Appendix B, Response 2) & \\

         ``The ledger rewrite has only ever been approved by my team when there's a governance proposal voted on by the blockchain to approve these specific operations.'' (Appendix B, Response 3) & \\
         ``Absolutely [we've run state-changing upgrades], and [we've run] more than I know about, I'm sure.'' (Appendix B, Response 4) & \\

         ``There's a reason why we will knowingly run code that alters a balance, and it's the same reason that we don't even check the source code. I think it's important to highlight it. Not only do we not have time to review the source code, but also frankly I don't think that even if we objected it would even matter.'' (Appendix B, Response 6) &\\

         \hline

    \end{tabular}
    \end{minipage}%
    \qquad\qquad\qquad\qquad
    \begin{minipage}[t]{.62\textwidth}
    \caption{Thematic analysis (b): 2 of 4}
    \label{table:results_coding2}
    \begin{tabular}{|p{0.8\linewidth}|>{\raggedright\arraybackslash}p{0.2\linewidth}|}
         \hline
         \raggedright \textbf{Concept or discrete example}\arraybackslash & \textbf{Theme} \\

         \hline
         ``We never quit a chain because we disagree with the community so far, and a lot of times we actually vote kinda according to the consensus too. We independently make the decision, our decision is aligned with the community, and that makes it easier. The tricky part would be if we disagree, and we voted the other way, but two thirds of the community still voted the other way, it's a tricky question.'' (Appendix B, Response 2) & Incentive Function \\

         ``I don't think that even if we found bad code, and escalated it, that anyone would do anything about it. Nor do I think that if we objected to ledger rewrites that it would even matter. If anything, validators that I know that have objected to ledger rewrites end up in more pain and trouble than if they just wouldn't have said anything. So I don't think we're incentivised to even try and protect the chain from that perspective.'' (Appendix B, Response 6) & \\

         ``We never check [upgrade code]. There's several rationales here. One is a time constraint, which is an interesting paradox. We have service level agreement s with our private clients where we have to maintain high degrees of uptime for these nodes\ldots A lot of our clients find that blockchain-based [downtime] sanctions are insufficient, so they require more blood in the water than the blockchain [downtime penalties] would provide.'' (Appendix C, Response 2) &\\

         ``When stake is low, like shitchains, even if nobody verifies the code, that's fine, it's just playing with magic internet money, but for high-stake chains like Solana, we know that some validators review it. Although, we never confirm it, so in a way, that's another decentralized consensus.'' (Appendix C, Response 5) &\\

         ``There are other chains where you have to vote to upload smart contracts onto their network, and what would happen is that most folks who are running the chain are making fifty bucks a week, a hundred if you're lucky. How much time does a person actually spend on that chain?\ldots Fixing things, making sure machines are up, thinking about the governance, trying to attract new people to delegate to you, and so on. How much time are you going to spend looking at software upgrades, or looking at the governance proposals that they have?'' (Appendix D, Response 1) &\\

         ``I think there are tonnes of people that should not be allowed to vote because they are completely irrational. Proof-of-stake and proof-of-work governance systems assume, in fact, I believe it's a theoretical underpinning, that all voters are rational market participants, and in crypto that is just not even true. People will vote for the most irrational reasons, people will sell their votes, people won't even vote.\ldots You should never have the pure [direct] democracy, and frankly, validators don't have a lot of time to be a representative. The validators that appear to be really good at being representative have just straight-up automated to vote yes on everything, so both are broken.'' (Appendix D, Response 6) &\\

         \hline

    \end{tabular}
    \end{minipage}
    \end{adjustbox}
\end{table}
}}
{\renewcommand{\arraystretch}{1.5}
\quad{\setlength{\tabcolsep}{1em}
\begin{table}[t!]
    \begin{adjustbox}{width={\textwidth},totalheight={\textheight},keepaspectratio}
    \begin{minipage}[t]{.62\textwidth}
    \caption{Thematic analysis (c): 3 of 4}
    \label{table:results_coding3}
    \begin{tabular}{|p{0.8\linewidth}|>{\raggedright\arraybackslash}p{0.2\linewidth}|}
         \hline
         \raggedright \textbf{Concept or discrete example}\arraybackslash & \textbf{Theme} \\

         \hline
         ``sometimes the source code is only released a few hours before the upgrade. In those cases, it might not even be possible to do a cursory look\ldots One problem is that some chains, even if they do go via [on-chain] governance for the upgrades, they might release the source code only close to the time that the proposal is to pass.'' (Appendix C, Response 1) &  Trust\\

         ``We would generally trust [code] implicitly, there are occasions where due to it being a security patch we don't have access to the code prior to the upgrade\ldots but it's mostly implicit trust on the Foundation team.'' (Appendix C, Response 3) &\\

         ``I think it is part of a validator's responsibility to be aware of what they're upgrading to. Again, it comes back to how confident you feel in the team that's running the upgrade, or running the chain'' (Appendix C, Response 4) &\\

         ``We don't verify the code, part of the reason is that we don't understand the code enough to do so ourselves. Part of the reason is that we don't believe that validators are supposed to do it. Some validators are able and capable of doing it, they do it, that's great, we give kudos to them.\ldots You could say, 'how are you one hundred per cent sure?' we are never sure. We rely on some kind of [social] consensus.'' (Appendix C, Response 5) &\\

         ``On emergency upgrades, when there's no other processes, I do check what it is that they're trying to do, or change. On the larger ones, never. I might look at the release notes, about what is supposed to be in there\ldots unless you're very familiar with the actual codebase, you're not going to even know this. You have to trust the central facility is working in the chain's best interest.'' (Appendix C, Response 6) &\\

         ``Like 90\% of the time we will review the code, to make sure that there is no sneaky update, if it is made available. If it wasn't made available, I would ask that it be made available, and then not apply the upgrade if they wouldn't make it available.'' (Appendix C, Response 7) &\\

         ``in general, if the risk was collusion by a Foundation, we can see if the Foundation voted, or if the individual voted.'' (Appendix D, Response 2) &\\
         
         \hline
    \end{tabular}
    \end{minipage}%
    \qquad\qquad\qquad\qquad
    \begin{minipage}[t]{.62\textwidth}
    \caption{Thematic analysis (d): 4 of 4}
    \label{table:results_coding4}
    \begin{tabular}{|p{0.8\linewidth}|>{\raggedright\arraybackslash}p{0.2\linewidth}|}
         \hline
         \raggedright \textbf{Concept or discrete example}\arraybackslash & \textbf{Theme} \\

         \hline
          ``at some point there was a vote to touch their account and grab tokens from it to secure the chain, so that one party wouldn't have such massive voting power. Well, because the chain should be immutable we were thinking that this should not be happening. [It was a] social attack, I would say, because it's all going through [on-chain] governance, and then if the majority decides something, then that's how it goes.'' (Appendix B, Response 1) &  Governance\\

          ``the community makes the decisions, and we ride along. It's very hard to pre-decide in the hard [cases]. But we are flexible. We don't hold to an absolute principle about the immutable nature of blockchain.'' (Appendix B, Response 2) &\\

          ``There's no proven way of doing governance, but it's good enough in a lot of cases.'' (Appendix D, Response 1) &\\

          ``Any sort of governance system is trying to get a group of individuals to work together and solve a problem, and basically create a compromise of some sort. Inevitably, in any of these systems, there's some inherent flaw of someone can take over, someone can sway the vote, or choose their particular positions, and the ones that I've seen in Cosmos, despite some flaws for sure, it seems like the least worst way of governing'' (Appendix D, Response 2) &\\

          ``The more complicated a governance system is, the more likely people can take advantage of it.'' (Appendix D, Response 3) &\\

          ``I personally hate Cosmos governance. Validators have a lot of say in the votes, but most of the proposals do not really concern validators at all.'' (Appendix D, Response 4) &\\

          ``The future of Bitcoin is essentially being decided by the core devs and the pool runners. Ethereum, even though there has been a good effort at decentralization, Vitalik is still very much not officially, but unofficially, the final word on where the software ends up going.'' (Appendix D, Response 5) &\\

          ``fully completely democratic governance systems are just not good\ldots there's a lot of fatigue, because every little thing has to have a governance vote, and people get tired of voting all the damn time, so people give up. There's just a lot of problems there. Then, fundamentally, on a lot of proof of stake chains, the validators are used as a sort of very undefined, androgynous form of a senator or representative, or something where the validators are supposed to act in more of a republic style governance system. I believe that both models are flawed.'' (Appendix D, Response 6) &\\

          ``I'm torn on it, because we do achieve a lot with governance, I think even the low-level way that Bitcoin handles governance, I think it works and it's an almost impossible thing to solve, governance at scale\ldots I think particularly with Cosmos, where everything is driven with [on-chain] governance, I think it has a tonne of problems, but at a low level we do achieve governance, and we do achieve community control, or at least community involvement with blockchains.'' (Appendix D, Response 7) &\\

         \hline
    \end{tabular}
    \end{minipage}
    \end{adjustbox}
\end{table}
}}

\section{Analysis}

\subsection{Trust and Legitimacy}
It is important to re-state the argument from our companion paper that in a system that relies on global shared state, only the manifestation of a rewrite changes, not the possibility that a rewrite will occur. In a situation where decision-making power is dispersed more widely, a large base of stakeholders or users can force change, while if decision-making power is concentrated, a single entity, such as a Foundation, or a small group, such as Foundation and developers, or large validators, can force change (Table \ref{table:leviathan_table_1}). Note also that \textit{Practical Immutability} is a higher bar than simply the absence of ``collapse'' in the Budish sense. In all the examples in Table \ref{table:rewrites}, which are referred to in our transcripts, the rewrites did not trigger ``collapse.'' To begin with, we will attempt to define \textit{Practical Immutability} as ``enough immutability to maintain trust in the governance of a network, from the perspective of agents in the network.'' It is situated at the intersection of agents' desire for true, ISO-style immutability on the one hand, and their acceptance of certain types of rewrite under certain situations. This is a lower bar of immutability than many definitions. Indeed, several public, permissionless ledgers maintain \textit{Practical Immutability} even in the presence of rewrites. This does however, represent a higher degree of trust than is often assumed of these so-called `low trust' networks. This tension is perfectly illustrated by the contradictory survey responses reproduced in the Appendix (Figures \ref{fig:q9} and \ref{fig:q12}), which show that \textit{fewer validators say they would alter a single account balance than have actually altered one}.

\begin{table}[t!]
    \centering
    {\renewcommand{\arraystretch}{1.5}
\quad{\setlength{\tabcolsep}{1em}
    \caption{Takeovers, coercion and collapse on permissionless ledgers}\label{table:leviathan_table_1}
    \begin{tabular}[t!]{|p{0.2\linewidth}|p{0.4\linewidth}|p{0.4\linewidth}|} \hline 
         \textbf{Dispersion of decision making power}&   \textbf{Permissionless ledger}\\ \hline 
         \textbf{High}&   \textbf{Takeover:} Users force change\\ \hline 
         \textbf{Medium}&  \textbf{Takeover:} Foundation, DAO, Validators, or large token holders force change\\ \hline 
         \textbf{Low}&   \textbf{Takeover:} Foundation or controlling DAO forces change\\ \hline
    \end{tabular}
    }}
\end{table}
\begin{table}[t!]
    \centering
    {\renewcommand{\arraystretch}{1.5}
\quad{\setlength{\tabcolsep}{1em}
    \caption{Example Ledger Rewrites in Cosmos and Ethereum}
    \begin{adjustbox}{width={\textwidth},totalheight={\textheight},keepaspectratio}
    \begin{tabular}[t!]{| p{0.2\linewidth} | p{0.5\linewidth} | p{0.3\linewidth} |}
         \hline
          \textbf{Incident}& \textbf{Description}& \textbf{Impact} \\
         \hline
         \raggedright Juno Proposal 16 & A targeted governance proposal to seize an entity's funds, valued at ~\$100M.~\cite{coindesvoteaway}~\cite{coindeskrevoketokens}~\cite{coindeskprop16watershed} & Funds were seized and locked in a smart contract. \\
         \hline
         \raggedright Osmosis clawbacks& Two governance proposals to mass alter balances and `claw back' unclaimed airdropped tokens.~\cite{mintscanosmosisprop29}~\cite{mintscanosmosisprop32} & Unclaimed airdropped OSMO and ION tokens were clawed back to the chain's Community Pool. \\
         \hline
         \raggedright Tombstone reversions & On the Chihuahua and Persistence networks, validators were tombstoned (a protocol-enforced penalization that permanently deletes both the node operator and delegators' tokens, and permanently removes that validator from the active set of nodes) for byzantine behaviour, and sought to have their penalization overturned via governance. ~\cite{mintscanpersistenceprop10}~\cite{mintscanhuahuaprop39} & The validators were re-instated, and ledger state altered to restore pre-slash balances. \\
         \hline
         \raggedright Ethereum post-DAO hack rewrite & After a hack, off-chain governance elected to rewrite the chain's state via hard-fork. & The chain forked, becoming two different chains, Ethereum and Ethereum Classic. \\
         \hline
    \end{tabular}
    \end{adjustbox}
    \label{table:rewrites}
    }}%
\end{table}
From our coding in Section 3, we identified five themes. From these five themes, we then collapse immutability, rewrites, and incentives (Tables \ref{table:results_coding1} and \ref{table:results_coding2}) into the category of \textit{Legitimacy} (which we theorize takes precedence over strict system integrity),\footnote{It is important that we argue stakeholders are seeking `legitimacy' of ledger state, rather than the more common `correctness.'} and trust and governance (Tables \ref{table:results_coding3} and \ref{table:results_coding4}) into \textit{Trust}, since the measure property we are interested in is trust \textit{in} the governance of a network. Thus we might supersede our initial definition, and switch the relationship between ledger state and its governance, instead stating that governance is the process by which trusted governance lends legitimacy to ledger state. 

\smallskip

\begin{center}
\begin{tikzcd}[column sep=5em]
    Agent \arrow{r}[]{Trust} & {Governance} \arrow{r}[]{Legitimacy} & {Ledger\:State}
\end{tikzcd}
\end{center}

\smallskip

We will describe in Section 4.2 that \textit{Trust} is both calculative and non-calculative, but for the moment we simply state its relationship to the likelihood that ledger state will be changed. This qualitative definition can be applied by analysing whether the relationship above holds, asking two simple questions: 

\begin{enumerate}
    \item Do agents \textit{trust} the network's \textit{governance topology}?
    \item Does the network's \textit{governance topology} see the ledger state\footnote{Depending on what is being examined, this could be current, or historical.} as \textit{legitimate}?
\end{enumerate}

Thus \textit{Practical Immutability} is simply the case where data is immutable, except where the data is deemed to be illegitimate. This explains the apparent supremacy of the governance topology of public permissionless blockchains over their technical network topology that we documented in our companion paper~\cite{lynham2025decentralizationqualitativesurveynode} and that is also implied by the work of Prasad et al.~\cite{quantifying_blockchain_immutability_over_time} on immutability. In fact, there is a symmetry with Bitcoin's motivation, where the design of the technology is framed in terms of trust, and solving the double-spend problem without an intermediary,~\cite{bitcoin_nakamoto} not the network-level trust dynamics we have discussed here.

\subsection{Trust}
A reasonable critique of this approach is that taken by De Filippi et al., which is to re-frame blockchains as ``Confidence Machines'', as a positive counter-argument to the negative connotations of `trust' (i.e. `risk'). ``Confidence does not entail personal vulnerability, because it emerges from prior experience or statistical evidence of how a system operates\ldots while trust is often associated with a feeling of uncertainty, confidence is generally associated with a feeling of predictability.''~\cite{DEFILIPPI2020101284} The outcome we are concerned with is legitimacy of state, which leads to conditional immutability; the outcome they are concerned with is predictability for agents, which leads to confidence. In that sense, these analyses may be complementary; in the prior flow diagram one could replace `trust' with `confidence' and the argument would hold. Where our analysis differs is we build on Williamson; they explicitly do not, and focus on actor confidence: ``Later scholars, such as Earle; have deemed calculativeness to be one of the attributes of a state of confidence.'' Ultimately however, their framing of confidence is not dissimilar to `institutional trust',

\blockquote
``confidence in a system ultimately depends on the level of trust or confidence that one has in the actors or institutions involved in higher-order systems. Giddens explains how much of the confidence we experience in our daily activities only subsists because of the trust we have in a variety of expert systems (e.g. the legal system, professional guilds, the scientific community, etc.) which we believe provide the necessary `guarantees' for us to build expectations on matters which we do not have the ability to exhaustively verify on our own. These expectations are grounded both on previous experience and common knowledge that these systems generally operate as expected, and on the trust assigned to a series of regulatory agencies responsible for overseeing these systems.''~\cite{DEFILIPPI2020101284}

\blockquoteend

Although it deals only tangentially with immutability, Curry's work on trust and legitimacy in blockchain bears mentioning. He classifies rewrites such as the Ethereum fork as ``self governance,'' observing, ``the Ethereum hard fork can be seen as an example of self-governance working, as the community majority voted for the fork. However, it still undermines the immutability of blockchain, and without effective decentralising mechanisms (which are hard to establish, as seen in the DAO case above), the removal of immutability can be problematic.''~\cite{10.1108/IJPSM-12-2023-0368} He notes that the bigger issue is not this form of governance but its limitations, ``with immutable decisions being reversed and hierarchies potentially reestablishing themselves in periods of crisis.''

His key argument is that the trust locus may rest elsewhere, with blockchains ``still rel[ying] on legitimacy conferred from other sources in order to be trusted by the public,'' by which he means the state, or regulation. Moreover, Curry sees legitimacy as a condition for the development of trust. We invert this relationship. Thus, trust is a precondition, not an output. Legitimacy of ledger state is instead the output, with governance legitimacy considered in a context of what De Filippi et al. term \textit{endogenous legitimacy}.~\cite{defilippi_blockchain_technology_legitimacy} We, to some extent, see trust and risk as table stakes in a system that deals in a primitive (cryptoassets) more akin to simplified credit than money. We intuit this based on the work on money versus simplified credit in Kahn et al.,~\cite{kahnetal} where defection of agents transacting using credit results in a return to autarky, or a preference for using a cash-like money. In this, we see a parallel to the discussion of token ``collapse'' by Budish et al.,~\cite{budish2024economiclimitspermissionlessconsensus} and ``undermin[ing] the system and the validity of [tokens]''~\cite{bitcoin_nakamoto} by Satoshi Nakamoto in Section 2.4. Ultimately, Curry's concern is ``not \textit{whether} blockchain creates trust or confidence, but how this trust is created (or not) by perceived legitimacy in blockchain governance.''~\cite{10.1108/IJPSM-12-2023-0368} In our conception, without trust, the endogenous token of a ledger would already have collapsed, since agents express themselves both calculatively and non-calculatively. 

An interesting question is whether or not there is a relationship between trust and decentralization. We have previously argued that greater decentralization does not necessarily mean greater resistance to rewrite.~\cite{lynham2025decentralizationqualitativesurveynode} Rather, what many are looking for in `greater decentralization' is, perhaps, greater trust in the governance of the system.\footnote{This may be at odds with a regulator; the implication in the MiCA wording is dispersion of authority, ``Where crypto-asset services are provided in a fully decentralised manner without any intermediary, they should not fall within the scope of this Regulation.''~\cite{mica} Arguably it is implicitly also robustness in the form of immutability.} The instances of user and stakeholder-driven rewrites show that diverse stakeholders seek to rewrite these networks, and that even rewrites that explicitly override the designed incentive mechanisms of these networks (for example, ``untombstoning,''---Appendix B, Response 1) can be borne with no ill effect, a fact that is not unintuitive to an observer. On the other hand, a rewrite that resulted in ``collapse''~\cite{budish2024economiclimitspermissionlessconsensus} of the network's endogenous token can also be done legitimately, so long as stakeholders decide the cost is worthwhile.

This is in accord with the ``hyphenated'' or ``institutional'' trust described by Williamson,

\blockquote
Although the environment is mainly taken as exogenous, calculativeness is not suspended but remains operative. That is because the need for transaction-specific safeguards (governance) varies systematically with the institutional environment within which transactions are located. Changes in the condition of the environment are therefore factored in---by adjusting transaction-specific governance in cost-effective ways. In effect, institutional environments that provide general purpose safeguards relieve the need for added transaction-specific supports. Accordingly, transactions that are viable in an institutional environment that provides strong safeguards may be nonviable in institutional environments that are weak---because it is not cost-effective for the parties to craft transaction-specific governance.\footnote{We conceptualize this as a check on opportunistic rewrites, afforded by the network topology, being subservient to the governance topology of that network.}~\cite{williamson1993}

\blockquoteend

Nearly all of Williamson's ``hyphenated'' trust examples are relevant to organisations with a topology like that found in blockchain networks (we enumerate these in Table \ref{table:williamson_hyphenated_trust}). Checking the arbitrary exercise of authority is key to political trust, and regulation can ``serve to infuse trading confidence into otherwise problematic trading relations.'' Though state regulation is inconsistent in the blockchain space, voluntary measures can arise organically if they are seen to boost trust, for ``regulation [can] have a (spontaneous) life of its own.''~\cite{williamson1993} Societal culture as a check on opportunism, and networks as a credible check on misbehaviour apply within an agent type (e.g. node operators) and between types (e.g. node operators and stakers), as we have discussed previously,~\cite{lynham2025decentralizationqualitativesurveynode} while the case of professionalization could be specifically applied to validators, who are expected to enact changes they do not agree with, ``The obligation to fulfil the definition of a role is especially important for professionals---physicians, lawyers, teachers, and so on. Although these roles generally arise in a spontaneous (evolutionary) manner, they are thereafter supported by entry limitations (such as licensing), specific ethical codes, added fiduciary obligations, and professional sanctions.''~\cite{williamson1993} These entry limitations and sanctions exist on public permissionless blockchains. Existing operators have access to informal back-channels, or private groups for upgrade notifications,\footnote{Such as those which were used to recruit respondents in the fieldwork for this paper.} while Foundation delegations and implied penalties for misbehaviour, as well as protocol-enforced ex-post penalties such as downtime slashing and tombstoning are similar to professional sanctions.

Together, these combine to describe the situation of conditional immutability where legitimacy and trust are key to the validity of any rewrite. Counterbalancing these is agents' assessment of risk, that is, calculative trust,~\cite{williamson1993} which may extend as far as a recognition of the risk of collapse as described by Budish et al.~\cite{budish2024economiclimitspermissionlessconsensus} Crucially, any rewrite that is seen as legitimate, particularly when viewed via the lens of a network's governance topology, does not appear to trigger the ``collapse'' referred to by Budish et al., the ex-post destruction in the value of an endogenous token that should accompany any attack on finality. This implies that from a qualitative definition, the rewrite was not in fact an attack at all.


Finally, we must observe that, although there is a tension in our fieldwork between a desire for immutability on the one hand but an acceptance of ledger rewrites on the other, it is a reasonable argument to instead say that immutability is not a design goal of these ledgers at all, and that they are best defined simply as mutable ledgers. Perhaps, to build on the analysis of Davidson et al., this is natural if blockchains are indeed ``a new institutional technology.''~\cite{DAVIDSON_DE_FILIPPI_POTTS_2018} 

Just as we agreed with the analysis of Vergne~\cite{vergne} previously that decentralization was on some level essentially performative, especially when conflated with immutability,~\cite{lynham2025decentralizationqualitativesurveynode} the reason perhaps is clearer from the tension just described. It follows that this trust-based definition of immutability is also to some degree performative. It relies on a trust in, perhaps even a belief in, the ``coherence of the play''~\cite{vergne} from all agents that participate in the system. It makes sense then, that those who confuse decentralization and immutability should experience both in a similarly performative way.
{\renewcommand{\arraystretch}{1.5}
\quad{\setlength{\tabcolsep}{1em}
\begin{table}[t!]
    \centering
    \caption{Summary of Williamson's Institutional, or `Hyphenated' trust types}
    \label{table:williamson_hyphenated_trust}
    \begin{adjustbox}{width={\textwidth},totalheight={\textheight},keepaspectratio}
    \begin{tabular}{|p{0.2\linewidth}|p{0.4\linewidth}|p{0.34\linewidth}|} 

         \hline
\raggedright \textbf{Hyphenated Trust}\arraybackslash &  \textbf{Summary}& \textbf{Potential relevance}\\
         \hline
         \raggedright A. Societal Culture& This refers to trust dynamics in large groups, for example a country. This trust ``involves very low levels of intentionality,'' as it is related to social norms. 
         & Applicable to the `culture' around a given network, e.g. expressed via stakers or users on social media. \\
         
         \hline
         \raggedright B. Politics & ``Legislative and judicial autonomy serve credibility purposes.'' This includes ensuring the application of laws; checks on arbitrariness, especially ``self-denying ordinances'' are important. & The autonomy and credibility of governance processes within a network's governance topology. The public actions of key stakeholders, like developers, Foundation members or validators are often heavily scrutinized by other agents. \\
         
         \hline
         \raggedright C. Regulation & ``[R]egulation can serve to infuse trading confidence into otherwise problematic trading relations.'' It can be spontaneous and voluntary. Both parties in a transaction should be able to conduct it on better terms than they would absent ``appropriate'' regulation.  & This is best typified by technical affordances like double-spend protection and smart contracts; however, a rewrite such as the one that followed the DAO hack on Ethereum might perhaps be classified as spontaneous regulation. \\
         
         \hline
         \raggedright D. Professionalization & Specialized roles typically involve responsibilities and obligations. Often, entry limitations are created, either informally or via professional bodies and certifications. & Developers and validators are held to informal standards. In governance, they are seen to have a representative responsibility. (See Appendix D, Response 6) \\
         
         \hline
         \raggedright E. Networks & Trading networks. ``[T]he maintenance of these networks depends on the perfection of intentional trading rules, [and] the enforcement of sanctions\ldots Credibility turns on whether these reputation effects work well or poorly.'' & Rules-based trading might best be typified by network topology features like double-spend protection, deterministic smart contract execution, or protocol-defined incentives. \\
         
         \hline
         \raggedright F. Corporate Culture & ``Corporate culture displays both spontaneous and intentional features and works mainly within particular organizations.'' This may be formal or informal; informal groups might form inside a formal organization. ``Internal effects spill over, moreover, onto external trade if firms take on distinctive trading reputations by reason of the corporate culture through which they come to be known and evaluated.'' & This is not only applicable to Foundations and Core Teams, but also describes agent interactions between categories and within categories, which may not be limited to a single chain (for example, validators that run across many - see Appendix F.4, Figs. 12, 14-17 in our companion paper).~\cite{lynham2025decentralizationqualitativesurveynode} \\ 
         
         \hline

    \end{tabular}
    \end{adjustbox}
\end{table}
}}

\section{Conclusion}

In this paper, we have examined the delta between a rigid definition of blockchain immutability, such as the one provided by ISO, and the reality of immutability reported by practitioners in our fieldwork. From aggregate themes found in our fieldwork of conditional immutability, rewrite events, incentive function, trust and chain governance, we have collapsed these to two meta-themes that contextualise the conditional nature of immutability on these networks, \textit{Trust} and \textit{Legitimacy}.

It is notable that validators and node operators mostly appear to respect the desires of chain governance even if the result is a controversial rewrite of ledger state. At a lower bar, even when validators have seen certain actions as illegitimate, they have apparently still kept enough trust in the governance of a network to decide not to exit. Perhaps this is a purely calculative step, or it might represent a more complicated interplay of motivations, perhaps framed within what Williamson calls ``Institutional trust.''~\cite{williamson1993}

In any case, this supports our general hypothesis that legitimacy is not directly conferred by agents in the system on the governance topology of a network. Instead, \textit{Trust} is conferred, and this trust allows the governance process of a network to be applied to ledger state to determine its \textit{Legitimacy}. We hypothesise that the visible manifestation of a collapse in trust in the governance of a network would be the same ``collapse'' described by Budish et al.,~\cite{budish2024economiclimitspermissionlessconsensus} or the reversion to autarky described in the simple credit models of Kahn et al.,~\cite{kahnetal} where agents cease transacting using credit (the credit in this case mapping to an endogenous token). Should this collapse occur, economic security would cease to function, resulting in the death of most existing designs of public, permissionless ledger.

That blockchains survive various types of rewrite, including targeted seizures, clawbacks, untombstonings and balance manipulation shows that rewrites are tolerated in certain cases. We argue this is a function of trust in governance lending legitimacy to ledger state. Even if an individual operator sees an action as illegitimate or even as an attack, they might still trust governance enough in the general case to decide not to exit. Alternatively, it might be a calculated decision as part of a `hyphenated' form of trust as described by Williamson, for example as a result of professionalization,~\cite{williamson1993} where validators understand it is key to their credibility to enact governance decisions even if they do not agree with them.\footnote{Indeed, several validators describe this tension in their responses in Appendix B.}

All of this leads to the conclusion that these ledgers are not designed to be immutable at all, merely proof against the most simple opportunistic actions by hostile agents. In such a system, careful maintenance of trust is required to avoid collapse. In this context, a governance-mandated rewrite is akin to a regulatory action as described by Williamson, re-establishing the legitimacy of ledger state. Indeed, this seems to be supported by certain validators, who imply a rewrite in response to an opportunistic action, such as a hack, would be justified.\footnote{See Appendix B, Response 2.}



\bibliographystyle{splncs04}
\bibliography{biblio}

\section{Appendix}

\appendix

In this Appendix are the relevant excerpts from our semi-structured interview transcripts.

\section{Semi-structured Interview discussion areas}

\begin{enumerate}[noitemsep]

\item How confident do you feel about the security and robustness of communications related to upgrades, security issues etc?
\item How often do you or your node operations team check the code contained in an upgrade before applying it?
\item Would you or your node operations team knowingly run code that caused a ledger rewrite that affected a single account balance?
\item Have you or your node operations team knowingly run code that caused a ledger rewrite that affected a single account balance?
\item What do you understand by the term ``decentralization''?
\item To what extent do you agree with the following statement: ``There is adequate decentralization on most of the networks we work on.''
\item What do you understand by the term ``decentralization theatre''?
\item Does your organisation validate or mine on any networks that are nominally competitors? What issues can arise?
\item Has your organisation ever experienced a conflict of interest between two networks your organisation validates or mines on?
\item Has your organisation run a node in the genesis set for a permissionless ledger?
\item How did your organisation get selected for that genesis set?
\item If your organisation has run a node in more than one, is there a general pattern for how your organisation has been selected?
\item As professional operators, how concerned are you about privacy on the public networks you work on?
\item What is your opinion on the level of maturity in blockchain governance?

\end{enumerate}

\section{Semi-structured Interviews - Ledger Rewrites}

These are responses from our semi-structured interviews to the question, ``Have you or your node operations team knowingly run code that caused a ledger rewrite?'' Note that the response numbers are arbitrarily assigned.

\subsection{Response 1}

That does happen. Sometimes there are clawbacks and sometimes there is untombstoning, and then there is [Juno] Prop 16. So it does happen. Clawbacks are when there has been for example an airdrop that needs to be claimed, and people have not claimed it. Then the unclaimed amount is clawed back into the [chain] community pool, for example. Untombstoning is cases when a validator has double-signed, for example, during an upgrade, and then they make a proposal to [get reinstated]. Tombstoning means that the validators are dead for life, basically, and untombstoning is where they try to undo the tombstoning caused by their double-sign mistake. That has happened a couple of times, and since this is a governance-based change, when the majority claims that this is fine, then basically everybody has to follow suit. In my opinion this should not be done. Whatever has happened on-chain, has happened, and it should not be changed afterward.

[Prompt: because a core assumption is immutability?] Yeah, that's the main reason for it. Doing rewrites of the state changes how these chains should be working.

[Prop 16] was a case on Juno network where one entity got airdropped, they had a lot of accounts on [the Cosmos Hub] and they got airdropped a massive amount of tokens, and then at some point there was a vote to touch their account and grab tokens from it to secure the chain, so that one party wouldn't have such massive voting power. Well, because the chain should be immutable we were thinking that this should not be happening. [It was a] social attack, I would say, because it's all going through [on-chain] governance, and then if the majority decides something, then that's how it goes.

\subsection{Response 2}

Yeah it happens. I don't know the percentages, but when it happens, typically people talk about it, because in a way it's controversial in the blockchain world. We think that it's an immutable ledger, but sometimes, shit happens, sometimes [there is a rewrite] for a good reason, like a hacker hacked some account, we want to take it away. Based on social consensus we say, you know, ``what he did is wrong, so we want to right the wrong.'' Sometimes it's more like, not even a hack, but ``is that person actually entitled to something?'' There you get into Juno proposal 16 kind of thing, but I would say that in general, it doesn't happen too often, and when it happens, typically there is a discussion. The community has a discussion about it and then different people make up their mind, and then [as a validator] our responsibility, once it's passed, we just upgrade the chain, right? If you disagree, you can always exit. It's permissionless.

[Prompt: does your organisation take a view on those rewrites? Is that the protocol working as intended?] Well, what is a protocol? It's kind of a nothing, unless it's social consensus. When an upgrade happens, somebody writes a proposal, and then they need two-thirds of VP, voting power, to pass. If it passes, that's social consensus. We don't have any issues upgrading a chain based on consensus. So we never really question it. We never quit a chain because we disagree with the community so far, and a lot of times we actually vote kinda according to the consensus too. We independently make the decision, our decision is aligned with the community, and that makes it easier. The tricky part would be if we disagree, and we voted the other way, but two thirds of the community still voted the other way, it's a tricky question. I don't know what we would do in that moment. Luckily so far it has not happened yet.

[Prompt: is immutability a core assumption of the technology?] We never take an absolutist approach. Everything has a limit. In America, they have the constitution, free speech, but you also have limits. You cannot threaten peoples' lives. So the same principle happens to us. In principle, blockchain is supposed to be immutable, but in some cases we carve out exceptions. Those exceptions will evolve. We know that if there is a hack, if it's something the protocol is not supposed to do, or somebody finds a vulnerability, if we can claw back the [affected] funds, we think that's the right thing to do. But that's an easy example. There are tricky examples, but we don't know, it's probably case law. As it goes, when a situation happens, the community makes the decisions, and we ride along. It's very hard to pre-decide in the hard [cases]. But we are flexible. We don't hold to an absolute principle about the immutable nature of blockchain.

[Prompt: so you were happy to run, for example the proposal 16 code provided it was a community decision?] Yeah.

\subsection{Response 3}

Only when approved by governance.

[Prompt: what is governance in this case? A proposal workflow, or off-chain consensus and a coordinated change by validators?] The ledger rewrite has only ever been approved by my team when there's a governance proposal voted on by the blockchain to approve these specific operations.

\subsection{Response 4}

Absolutely [we've run state-changing upgrades], and [we've run] more than I know about, I'm sure. I don't know the ins and outs of every single upgrade that we've ever run. I know that they do change state, and I know that in some cases, they don't even communicate that they are changing state.

\subsection{Response 5}

Not on any production mainnets, only on test networks.

\subsection{Response 6}

Oh yeah, totally. Specifically in regards to the Juno Whale incident. Of course [we've run clawbacks], there's a blockchain that we already did that on. There's a reason why we will knowingly run code that alters a balance, and it's the same reason that we don't even check the source code. I think it's important to highlight it. Not only do we not have time to review the source code, but also frankly I don't think that even if we objected, that it would even matter. So I have a very nihilistic perspective on providing feedback. I don't think that even if we found bad code, and escalated it, that anyone would do anything about it. Nor do I think that if we objected to ledger rewrites that it would even matter. If anything, validators that I know that have objected to ledger rewrites end up in more pain and trouble than if they just wouldn't have said anything. So I don't think we're incentivised to even try and protect the chain from that perspective. There's a legal liability component where if you don't take a view, at least you're out of the spotlight, but if you're out there, arguing in the public domain one way or the other, you may become a target for a civil lawsuit. Like, some random user of a blockchain may attempt to sue you. I do know multiple validators that have had to quote-unquote lawyer up in these types of situations, where if they had just shut up and kept their head down, they probably wouldn't have had to. [On chain voting and off-chain discourse] can turn you into a target.

\section{Semi-structured Interviews - Upgrades}

The following are responses from our semi-structured interviews to the topic area, ``How often do you or your node operations team check the code contained in an upgrade before applying it?'' Note that the response numbers are arbitrarily assigned.

\subsection{Response 1}

I try to check the code, but there's a lot of problems with that. One problem is that it's quite common that it's only the inclusion of new versions of dependencies. It is next to impossible to check everything, especially when there's a lot of changes. Also, sometimes the source code is only released a few hours before the upgrade. In those cases, it might not even be possible to do a cursory look.

[Prompt: is that disorganisation, or for security reasons?] It is a bit of both, so that's very common for security upgrades, but those typically are a lot smaller, so you might get some idea of the code in a short period of time if you happen to have time to review it when it's released. One problem is that some chains, even if they do go via [on-chain] governance for the upgrades, they might release the source code only close to the time that the proposal is to pass. I think it is more about the way they work, rather than about being disorganised. But it could also be that they are not up to the task.

\subsection{Response 2} 

Never. Absolutely never. We never check. There's several rationales here. One is a time constraint, which is an interesting paradox. We have service level agreements with our private clients where we have to maintain high degrees of uptime for these nodes. So if we dilly dally, and take too long to do an upgrade, there's a high probability that node will be offline. The way that we structure our service level agreements is that we are subject to economic penalties for extreme forms, or even moderate forms of downtime. These are not [protocol enforced], these are direct contractual obligations between our company and our clients. These are over-the-counter private penalties. A lot of our clients find that blockchain-based [downtime] sanctions are insufficient, so they require more blood in the water than the blockchain [downtime penalties] would provide. That's across the board [on proof-of-stake chains].

[Prompt: do you think that the economic incentives around proof-of-stake work?] No, I don't. I think that proof-of-stake was a great idea, but it definitely does not work. There are, I think, more down sides to proof-of-stake than up. I think proof-of-work is a bit silly, and I think proof-of-stake is a bit silly. Do I have a constructive offering for a third or a fourth protocol? No I don't, but both of them I think are fundamentally flawed.

\subsection{Response 3}

We would generally trust implicitly, there are occasions where due to it being a security patch we don't have access to the code prior to the upgrade. If the code is available before the upgrade, we will occasionally review, but it's mostly implicit trust on the Foundation team.

\subsection{Response 4}

I think it is part of a validator's responsibility to be aware of what they're upgrading to. Again, it comes back to how confident you feel in the team that's running the upgrade, or running the chain, for example, and publishing the upgrades. I have looked over chain upgrades before, gone over the code and made sure that I have at least a vague understanding, but validators generally aren't developers, to the extent that they're involved in the chain code, and actually, digging into the chain code and understanding all of the edge cases and all of the cases where something can happen, or something malicious could be snuck into it, is virtually impossible. I think in the validator community as a whole, there should be people doing that, but I don't think it falls at the validators' feet as the people who should be catching that sort of thing. I think it should go hand-in-hand with regular audits from the chains, again, depending on how much value is actually at risk there. I think that validators are a last resort in terms of checking chain code.

I think that if everybody just thinks ``the validators are checking it,'' that's a very dangerous place to be, because I think it's very rare that people are checking it to at least the level that it would need to be, in order to be confident about it.

\subsection{Response 5}

We don't verify the code, part of the reason is that we don't understand the code enough to do so ourselves. Part of the reason is that we don't believe that validators are supposed to do it. Some validators are able and capable of doing it, they do it, that's great, we give kudos to them. I don't think that's a validator's responsibility. Especially if you have a decentralized network with a thousand validators. Then you can rely on sampling. If you say ten percent or five percent of validators can do it, if we are confident enough that is happening, then [as] a validator [we] can move on, without reviewing the code. That's what I believe. Especially for us, we are a very general validator in different ecosystems. It's almost impossible to specialise on all the code. If we did, we wouldn't be any good at any of those anyway. We wouldn't be good at reviewing the code, so we typically rely on some sort of common sense. When stake is low, like shitchains, even if nobody verifies the code, that's fine, it's just playing with magic internet money, but for high-stake chains like Solana, we know that some validators review it. Although, we never confirm it, so in a way, that's another decentralized consensus. We know some people verify it once the code has been released for a few days, we know the code has been checked, so we just move on. You could say, 'how are you one hundred per cent sure?' we are never sure. We rely on some kind of [social] consensus.

[Prompt: is there ever evidence or proof of this checking?] Not to my knowledge. It's the sort of trust that we always do in every day life. If you walk into the street, your assumption is that people will stop, the cars will stop. Of course, you could say, ``how do you one hundred per cent trust them? Have you asked them? You don't even know who the next driver is.'' You kind of rely on this common sense. The same principle applies here.

\subsection{Response 6}

On emergency upgrades, when there's no other processes, I do check what it is that they're trying to do, or change. On the larger ones, never. I might look at the release notes, about what is supposed to be in there. The problem is, that especially in Cosmos, there's no one spot to see all the different things that have changed, because they're all in submodules and all these other different things that they're relying on. Typically with security patches what happens is your main piece of code to add is changing a version number or one or two lines, but you're upgrading a package, that then upgrades another package, and that's where all the code has been changed. So unless you're very familiar with the actual codebase, you're not going to even know this. You have to trust the central facility is working in the chain's best interest. 99.999\% of the time, they are, but where a validator has seen some issues is where you have a tired engineer, he upgrades something, and then puts a bug back into the system. So there's a bugfix on a Cosmos security release, but because he was working on a different version of the [git] tree, we patched it at the time, and maybe a month later, his new version didn't have that patch, and he applied it. I don't think anybody would have caught it, but this is where some of the control processes break down. The code is just too big, and there's not many people that look at those things. Everyone relies on somebody else. A lot of validators don't have a software background, so I would presume that they check zero. A lot of validators are great at doing devops, which is a certain skill, versus software engineering, versus knowledge of the blockchain internals. There aren't that many people that know the blockchain and can write software modules. You might be a smart contract developer but you don't know all the ins and outs of what's going on in a blockchain, it's pretty esoteric stuff. If I had to guess a number, I'd say that ninety percent don't.

[Prompt: Is it an assumption that validators check code before they run it?] That's what the book says! They're the ones that look at the upgrades, look at the release notes, see what's changed, and theoretically make sure that processes are being followed. That's what you're supposed to do.

\subsection{Response 7}

Like 90\% of the time we will review the code, to make sure that there is no sneaky update, if it is made available. If it wasn't made available, I would ask that it be made available, and then not apply the upgrade if they wouldn't make it available.

\section{Semi-structured Interviews - Governance Maturity}

The following are responses from our semi-structured interviews to the topic area, ``What is your opinion on the level of maturity in blockchain governance?'' Note that the response numbers are arbitrarily assigned.

\subsection{Response 1}

It depends on the chain, and every chain is different. There is no book that new Foundations, or new chain people get, saying ``ah, you want to create a chain? Here is the elite way of doing governance.'' They all bumble along and do it. Some of them, they do the basic mechanics, so most Cosmos chains, when they're doing upgrades, they're doing them properly. As in, via the [on chain governance] upgrade notification. Some chains still don't. It depends on the chain. There's no proven way of doing governance, but it's good enough, in a lot of cases. For me, the personal irritation is where you do a governance vote, and it lasts two hours. I think Canto was a perfect example of this. They would have a governance vote that lasted two hours. So if you were asleep, you missed it. There are other ones that last seven days. Seven days is way too long, especially if you're trying to move the market, or things are going on. That's one example, of one of the parameters. Every time you do one of these and change one of the parameters, there's a whole load of other systemic effects which come through in game theory about what can happen, so there's no perfect [protocol]. There are other chains where you have to vote to upload smart contracts onto their network, and what would happen is that most folks who are running the chain are making fifty bucks a week, a hundred if you're lucky. How much time does a person actually spend on that chain? If you're making a hundred dollars a week on that chain, well you need to make three thousand a week in order to [survive], so a hundred dollars divided by three thousand is how much time you get to spend on that chain. Fixing things, making sure machines are up, thinking about the governance, trying to attract new people to delegate to you, and so on. How much time are you going to spend looking at software upgrades, or looking at the governance proposals that they have?

\subsection{Response 2}

Any sort of governance system is trying to get a group of individuals to work together and solve a problem, and basically create a compromise of some sort. Inevitably, in any of these systems, there's some inherent flaw of someone can take over, someone can sway the vote, or choose their particular positions, and the ones that I've seen in Cosmos, despite some flaws for sure, it seems like the least worst way of governing what could be any level of shady actors or collusion or interest, because the history is there and you can work back from that history to identify where someone was dishonest or maybe trying to bend the rules. Now, some of this can be clandestine and impossible to truly reveal, but as the current state of the art, it's not bad. It's not great, it's not perfect, but I would say in sort of a crawl, walk, run, we're probably just at the inside of crawl, and maybe approaching walk. Certainly there's tonnes of improvements that could be made to make things more fair, or what have you, but in general, if the risk was collusion by a Foundation, we can see if the Foundation voted, or if the individual voted, and then comes the question, ``well, was the individual an actual individual?'' It's hard to say, but so far, you know.

[Prompt: how can you manage the risk that the pressure is exerted outside of any visible channel?] Well, the challenging part of course is the anonymity that's inherent with these networks, by not being able to attach the identity of the voter to any real entity you have to blindly trust that they are who they say they are, but inherently you can't verify that. So it is a challenge. Maybe an impossible one.

\subsection{Response 3}

I think it's pretty good. I know there's a lot of people that are trying to improve governance. For me, it should be the opposite. I feel that we overthink the governance too much, and sometimes governance is really hard to improve. If you look at Polkadot, they have Governance 2.0, which is supposed to solve all the problems. But it actually introduces a lot of new problems.

[Prompt: what would be an example of something they tried to solve that became more complicated?] They made different tracks. In Polkadot, before, when you did spending it was just one type of proposal, and now I believe there's a big spender, medium spender, high tip, low tip, and each of them have a different threshold. It sounds great, you're making it more nuanced, but it turns out to be a mess. First, people don't understand it, and then more people do not participate, and then you further increase the chance of whales swinging the vote. The more complicated a governance system is, the more likely people can take advantage of it. So that's my point of view. You can say that's wrong or right, but that's how I see it. I think too many people focus too much on governance, before we even have a viable product. Blockchain is still early, I feel like we need to focus on other things before we talk about governance.

[Prompt: like what?] Develop a product that people want to use, rather than debating something philosophically. Are millions of people using the product today? People sometimes take the wrong approach. People like stablecoins, but purists will say ``this is wrong,'' well, who are you to tell me it's wrong? Same for governance. If nobody cares, then just stop doing it. Just develop the product to where more people use it. The more people use it, that increases the value of the chain, then the time will come to develop some new infrastructures in governance to manage that multi-billion dollar blockchain. So many chains take the opposite approach. They first want to perfect governance, and DAOs, before anybody is using the chain. What's the point?

\subsection{Response 4}

I personally hate Cosmos governance. Validators have a lot of say in the votes, but most of the proposals do not really concern validators at all. For example, if we are talking about community pool spends, then the say should be from people who actually know about economics. The validator business by itself is typically for technical people and everybody has a say in technical matters, financial matters, and so forth, even though they have no clue about it. Whenever there is a controversial vote then it is going to be a massive influencing campaign by some vocal people to affect the vote, and it doesn't make sense. It's not proper governance. Nothing is optimal, how it is currently done in Cosmos.

\subsection{Response 5}

We're in the very early days. The future of Bitcoin is essentially being decided by the core devs and the pool runners. Ethereum, even though there has been a good effort at decentralization, Vitalik is still very much not officially, but unofficially, the final word on where the software ends up going. I've played around on a lot of technologies built on top of things like Ethereum, like Decentralized Autonomous Organisations, even those kind of fall short. While the idea of decentralizing governance is fantastic, the people who create these DAOs, or products like Aptos, no-one actually wants to cede power, so I've never seen one work in a decentralized way.

\subsection{Response 6}

I think blockchain governance is very immature, and I think that most of it has to do with the architecture of proof-of-stake. I have very limited faith in proof-of-stake, and the effectiveness of proof-of-stake style governance. I think it's fundamentally flawed. Perhaps this is a larger philosophical statement but I believe that fully, and I can't believe I'm saying this, but I'm straight-up saying this, fully completely democratic governance systems are just not good. There was a Game of Thrones scene where the nobles were sitting there laughing about how dumb the idea of democracy is; at the time I was disgusted by that concept, but after doing this in crypto, I completely agree. I think there are tonnes of people that should not be allowed to vote because they are completely irrational. Proof-of-stake and proof-of-work governance systems assume, in fact, I believe it's a theoretical underpinning, that all voters are rational market participants, and in crypto that is just not even true. People will vote for the most irrational reasons, people will sell their votes, people won't even vote. I think there's a lot of fatigue, because every little thing has to have a governance vote, and people get tired of voting all the damn time, so people give up. There's just a lot of problems there. Then, fundamentally, on a lot of proof of stake chains, the validators are used as a sort of very undefined, androgynous form of a senator or representative, or something where the validators are supposed to act in more of a republic style governance system. I believe that both models are flawed. You should never have the pure [direct] democracy, and frankly, validators don't have a lot of time to be a representative. The validators that appear to be really good at being representative have just straight-up automated to vote yes on everything, so both are broken.

\subsection{Response 7}

I'm torn on it, because we do achieve a lot with governance, I think even the low-level way that Bitcoin handles governance, I think it works and it's an almost impossible thing to solve, governance at scale. So I think it is achieved and we do quite well with it. Whether I'd say it's mature, or whether crypto as a whole is anything close to mature, I don't know, but I think particularly with Cosmos, where everything is driven with [on-chain] governance, I think it has a tonne of problems, but at a low level we do achieve governance, and we do achieve community control, or at least community involvement with blockchains. I think there's still a long way to go, but I'm on balance impressed with how much involvement there is from the community in decision-making.

\section{Survey responses}

Here we have replicated the figures that illustrate responses to questions 9-12 of our survey. The full responses can be found in our companion paper.~\cite{lynham2025decentralizationqualitativesurveynode}

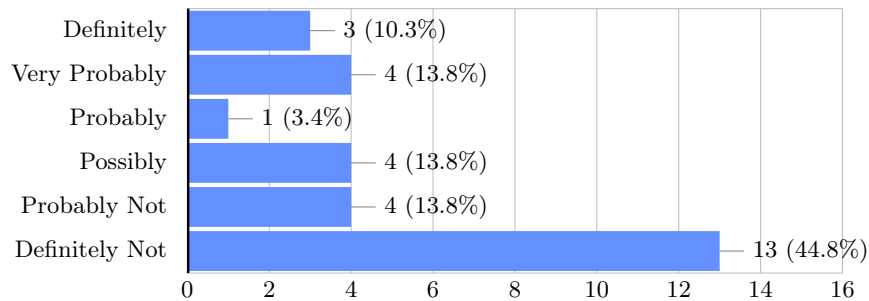
\begin{figure*}[ht]
\centering
\begin{tikzpicture}
\begin{axis}[%
width=0.84\textwidth, height=2in,
xbar, bar width=15pt,
xmin=0, xmax=16,
  symbolic y coords={dn,pn,poss,prob,vp,def},
  ytick=data,
  yticklabels={
    {Definitely Not},
    {Probably Not},
    {Possibly},
    {Probably},
    {Very Probably},
    {Definitely}},
y tick label style={align=right,text width=3cm},
xmajorgrids,
axis line style={lightgray},
major tick style={draw=none},
nodes near coords,
point meta=explicit symbolic,
node near coords style={font=\footnotesize,right=1em,pin={[pin distance=1em]180:}}
]
\addplot [
  fill={ibmblue},draw=none] 
  coordinates {
    (13,dn) [13 (44.8\%)]
    (4,pn) [4 (13.8\%)]
    (4,poss) [4 (13.8\%)]
    (1,prob) [1 (3.4\%)]
    (4,vp) [4 (13.8\%)]
    (3,def) [3 (10.3\%)]
  };
\draw [line width=1.5pt] (current axis.south west) -- (current axis.north west);
\end{axis}
\end{tikzpicture}
\caption{Q12. In the event of a hard slash that affected a validator run by your organisation, how likely is it that you would seek to undo the effect on the validator and delegators (e.g. via governance)?}
\label{fig:q12}
\end{figure*}

\section{Thematic Analysis Intermediate Coding}

{\renewcommand{\arraystretch}{1.8}
\quad{\setlength{\tabcolsep}{1em}
\begin{table}[ht!]
    \centering
    \caption{Thematic analysis intermediate coding}
    \label{table:coding}
    \begin{adjustbox}{width={\textwidth},totalheight={\textheight},keepaspectratio}
    \begin{tabular}{|p{0.8\linewidth}|>{\raggedright\arraybackslash}p{0.2\linewidth}|}
         \hline
        \raggedright \textbf{Intermediate Theme}\arraybackslash & \parbox{0.2\textwidth}{\centering\textbf{Aggregate Theme}}\medskip \\

         \hline
         Strict adherence to principle of immutability & \multirow{3}{*}{\parbox{0.2\textwidth}{\centering Conditional Immutability}} \\

         Willingness to modify state &\\

         Validator verification of state and upgrades &\\

         \hline
         Untombstoning events & \multirow{6}{*}{\parbox{0.2\textwidth}{\centering Rewrite Events}} \\

         Clawback events &\\

         Governance-driven rewrites &\\

         Targeted seizures &\\

         Validators enacting governance decisions &\\

         On-chain governance &\\

         \hline
         Validator incentives & \multirow{6}{*}{\parbox{0.2\textwidth}{\centering Incentive Function}} \\

         Staker incentives & \\

         Risk of token collapse &\\

         Risk of cyberattack via code or upgrade &\\

         Governance fatigue &\\

         Voter rationality or calculativeness &\\

         \hline
         Foundation or Core Team centralization & \multirow{4}{*}{\parbox{0.2\textwidth}{\centering Trust}} \\

         Centrality of validator set &\\

         Centralization of emergency upgrades &\\

         Centralization of code review and due diligence &\\

         \hline
         Governance supremacy over network topology & \multirow{6}{*}{\parbox{0.2\textwidth}{\centering Governance}} \\

         Unpredictability of governance votes &\\

         Centralization of governance power &\\

         Delegated governance responsibility &\\

         On-chain governance &\\

         Off-chain governance &\\
         
         \hline

    \end{tabular}
    \end{adjustbox}
\end{table}
}}



%
%
%

\end{document}